\documentclass[12pt,a4paper]{article}

\usepackage{epsfig}

\begin{document}
\vspace*{-3cm}
\begin{flushright}
FISIST/08--2002/CFIF \\
UG--FT--141/02, CAFPE--11/02 \\
hep-ph/0208171 \\
August 2002
\end{flushright}
\vspace{0.5cm}
\begin{center}
\begin{Large}
{\bf Precise determination of the $Wtb$ couplings at LHC}
\end{Large}

\vspace{0.5cm}
F. del \'Aguila \\
{\em Departamento de F\'{\i}sica Te\'{o}rica y del Cosmos \\  
Universidad de Granada, E-18071 Granada, Spain} \\
\vspace{0.5cm}
J. A. Aguilar--Saavedra \\
{\em Departamento de F\'{\i}sica and CFIF \\
Instituto Superior T\'ecnico,
P-1049-001 Lisboa, Portugal}
\end{center}

\begin{abstract}
Top pair production at LHC is the ideal place to search for nonstandard $Wtb$
couplings in $t \to W b \to l \nu b$ decays. The $lb$ forward-backward
asymmetry in the $W$ rest frame is very sensitive to $\sigma^{\mu \nu}$
couplings, and
can spot one-loop QCD corrections to the decay vertex with more than $5\sigma$
statistical significance. We discuss the potential of this asymmetry to signal
nonstandard $\gamma^{\mu}$ and $\sigma^{\mu \nu}$
couplings and compare with top-antitop spin correlation
asymmetries, which have a lower sensitivity. We also
briefly summarise the results for Tevatron.
\end{abstract}


\section{Introduction}
LHC will be by far the largest source of
top quarks available in the forthcoming years, with a top pair cross-section
of 860 pb \cite{papiro1} and a total single top (plus antitop) cross-section of
306 pb from three processes \cite{papiro2,papiro2b,papiro2c}. This will allow to
perform precision studies of top couplings. In the Standard Model (SM) the
$Wtb$ vertex is purely left-handed and its size is given by the
Cabibbo-Kobayashi-Maskawa (CKM) matrix element $V_{tb}$. Unfortunately,
the $t \bar t$ cross-section is rather insensitive to its actual value,
assuming that it is much larger than $V_{td}$ and $V_{ts}$.
Hence, to obtain a measure of the {\em absolute} value
of $V_{tb}$ it is necessary to fall back on less abundant single top
production \cite{papiro2}, with a
rate proportional to $|V_{tb}|^2$. 
Still, $t\bar t$ production can give invaluable information on the $Wtb$
vertex. Angular asymmetries between decay products are very sensitive to
a small admixture of a right-handed $\gamma^{\mu}$ term
or a $\sigma^{\mu \nu}$ coupling of either chirality.
That is, if we parameterise the most general CP-conserving
$Wtb$ vertex with the effective Lagrangian
\footnote{The most general $Wtb$ vertex (up to dimension five) involves ten
operators, but at the level of precision of these asymmetries it is an excellent
approximation to consider the top on-shell. With $b$ also on-shell and
$W \to l \nu,jj$ six of them can be eliminated using Gordon identities. The
resulting Lagrangian can be further restricted assuming CP conservation. The
couplings can then be taken to be real, of either sign.}
\begin{eqnarray}
\mathcal{L} & = & - \frac{g}{\sqrt 2} \bar b \, \gamma^{\mu} \left( V_{tb}^{L}
P_L + V_{tb}^R P_R
\right) t\; W_\mu^- \nonumber \\
& & - \frac{g}{\sqrt 2} \bar b \, \frac{i \sigma^{\mu \nu} q_\nu}{M_W}
\left( g^L P_L + g^R P_R \right) t\; W_\mu^- + \mathrm{h.c.} \,,
\label{ec:1}
\end{eqnarray}
these asymmetries are sensitive to the values of $V_{tb}^R$, $g^L$ and
$g^R$ {\em relative} to the SM coupling $V_{tb}^L \equiv V_{tb}$,
which from now on will be normalised to
one. These new couplings vanish at tree-level in the SM, but
can be generated at higher orders in the SM or its extensions \cite{papiro1}.
In particular,
one-loop QCD corrections prompt the appearance of
a coupling $g^R = -0.00642$ \cite{papiro3}
that is detectable at LHC, as we argue in the following.

It is worth drawing attention to the importance of a direct measurement
of these parameters. Unless there is some accidental cancellation among
new contributions, new physics scenarios are bound to bring about
the appearance of new observable $V_{tb}^R$, $g^L$ or $g^R$ terms, which are
likely to be comparable in size to the one-loop QCD correction to $g^R$ above.
It has
been shown that in supersymmetric or two Higgs doublet models radiative
corrections can give enhanced contributions to the top width 
\cite{papiro3b,papiro3c},
but these corrections should better manifest themselves in
angular asymmetries, whose study can be carried out with high precision
at LHC. On the other hand, low energy physics measurements do not set
model-independent constraints on these parameters.
Usually $V_{tb}^R$ is neglected in the
literature arguing that the presence of
such coupling at a detectable level would increase
the $b \to s \gamma$ branching ratio, resulting in
a value larger than the measured rate $\mathrm{Br} (b \to s
\gamma) = 3.15 \times 10^{-4}$ \cite{papiro5}.
(The $\sigma^{\mu \nu}$ terms are not discarded with this argument because they contain an
extra $q_\nu$ factor suppressing their contribution for small $q_\nu$.)
However, this is not compulsory since the amplitude involving $V_{tb}^R$ contains
the product $V_{ts} V_{tb}^R$, and $V_{ts}$ is not directly measured \cite{papiro5b}.
The SM value $V_{ts} \simeq 0.04$ is obtained assuming $3 \times 3$ CKM
unitarity, that does not hold if heavier fermions exist. 
So, in principle the value of $V_{ts}$ can be a fraction of its SM
estimate, and the indirect limit $|V_{tb}^R| \leq 0.04$ \cite{papiro4}
can be relaxed without spoiling the prediction for $b\to s \gamma$ and
other processes \cite{papiro6}.

At LHC these nonstandard couplings can be measured in single top production
\cite{papiro7,papiro21},
being the expected $2\sigma$ limits (with a realistic assumption of 5\%
systematic uncertainties) $-0.052 \leq g^L
\leq 0.097$, $-0.12 \leq g^R \leq 0.13$ \cite{papiro7}. In this Letter we
show that these bounds can be further improved in top pair production
$pp \to t \bar t \to WbW\bar b \to l\nu\!jjjj$
with the analysis of the $lb$ forward-backward (FB)
asymmetry in $t \to l\nu b$ decays, first proposed in \cite{papiro8}.
Not only it is more sensitive than single top production but it also
has smaller systematic uncertainties and does not depend on the details of the
production process. Here we investigate the
dependence of this asymmetry on all the anomalous couplings
in Eq.~(\ref{ec:1}), comparing its sensitivity
with that of other popular spin correlation asymmetries \cite{papiro1}.
The main background from $W\!jjjj$ is taken into account, and we perform a
simple simulation of the detector effects.

\section{Asymmetries in $t \bar t$ decays}

The FB asymmetry in the decay of the top quark $t \to W^+ b \to l^+ \nu b$
is defined as
\begin{equation}
A_\mathrm{FB} = \frac{N(x_{bl} > 0) -
N(x_{bl} < 0)}{N(x_{bl} > 0) +
N(x_{bl} < 0)}\,,
\end{equation}
where $x_{bl}$ is the cosine of the
angle between the 3-momenta of the $b$ quark and the
charged lepton in the $W$ rest frame, and $N$
stands for the number of events.
The same definition holds for the $\bar t \to l^- \bar \nu \bar b$ decay.

Angular asymmetries involving the
top spin rely on the fact that
top pairs are produced with correlated spins, decaying before
depolarisation takes place.
In order to fix the notation
we briefly summarise how angular asymmetries can be built from
spin correlations (see for instance Ref.~\cite{papiro1}).
A complete set of spin correlation observables for $t \bar t$ production
is described in Ref. \cite{papiro13}.
Let $\hat \mathbf{p}_t$, $\hat \mathbf{p}_{\bar t}$ be the normalised top and
antitop 3-momenta in the CM frame, and $\mathbf{s}_t$, $\mathbf{s}_{\bar t}$
their spins. We will study angular asymmetries based on the observables
$(\hat \mathbf{p}_t \cdot \mathbf{s}_t) \,
(\hat \mathbf{p}_{\bar t} \cdot \mathbf{s}_{\bar t})$,
which provides the spin correlation in the helicity basis, and
$(\mathbf{s}_t \cdot \mathbf{s}_{\bar t})$.
We do not consider the beam line or mixed bases, since the helicity basis
exhibits the highest degree of correlation at LHC \cite{papiro14}.
To build these angular asymmetries
the spins $\mathbf{s}_t$, $\mathbf{s}_{\bar t}$ can be inferred from the
distribution of the decay products. For a left-handed $Wtb$ vertex the
angular distribution of the fermion $f$ with respect
to the top spin in the top
rest frame is given by \cite{papiro14b}
\begin{equation}
\label{ec:3}
\frac{d\Gamma}{d \cos \theta_f} =
\frac{1}{2} \left( 1+h_f \cos \theta_f \right) \,,
\end{equation}
where $\theta_f$ is the angle between the fermion 3-momentum in the top rest
frame $\mathbf{p}_f$ and the top spin $\mathbf{s}_t$,
and $h_f$ are constants between $-1$ and $1$. For the antitop quark, the
distributions are obtained from the above formula with similar definitions
but replacing $\cos \theta_f$
by $-\cos \theta_f$.
For charged leptons and $d,s$ quarks $h_f = 1$ and the correlation
is maximal.
For neutrinos and $u,c$ quarks $h_f = -0.31$.
For leptonic decays we select
the charged lepton as spin analyser. For hadronic decays, the best choice
would be to select the down-type jet. However, the $d,s$ jets cannot be
experimentally identified,
and they are conventionally assigned to
the jet from the $W$ decay with smaller energy
in the top rest frame \cite{papiro16}. This
corresponds to a $d$ or $s$ 61\% of the time, and has an average correlation
with the top spin $h_q = 0.51$. Corresponding to these spin correlations,
we build the asymmetries
\begin{equation}
A_1 = \frac{N(x_l x_q > 0) - N(x_l x_q < 0)}{N(x_l x_q > 0) + N(x_l x_q < 0)}\,,
\label{ec:4}
\end{equation}
\begin{equation}
A_2 = \frac{N(x_{lq} > 0) - N(x_{lq} < 0)}{N(x_{lq} > 0) + N(x_{lq} < 0)}\,,
\label{ec:5}
\end{equation}
where $x_l$ is the cosine of the angle between $\mathbf{p}_l$
and $\hat \mathbf{p}_t$, $x_q$ the analogue for $\mathbf{p}_q$
and $\hat \mathbf{p}_{\bar t}$, and $x_{lq}$ the angle between $\mathbf{p}_l$
and $\mathbf{p}_q$.
Besides, we consider the correlation
$(\hat \mathbf{p}_t \cdot \mathbf{s}_t) \,
(\hat \mathbf{p}_{\bar t} \cdot \mathbf{s}_{\bar t})$ using the neutrino
as spin analyser in semileptonic decays 
\cite{papiro15}. This yields the asymmetry
\begin{equation}
A_3 = \frac{N(x_\nu x_q > 0) - N(x_\nu x_q < 0)}{N(x_\nu x_q > 0) +
N(x_\nu x_q < 0)} \,,
\label{ec:6}
\end{equation}
with $x_\nu$ the cosine of the angle between $\mathbf{p}_\nu$ and
$\hat \mathbf{p}_t$.
When nonstandard couplings are present, the angular distribution of the
decay products no longer corresponds to Eq.~(\ref{ec:3}).
Therefore, comparing the measured
angular asymmetries with their SM expectations
the presence of anomalous $Wtb$ couplings
can be established. However, these
asymmetries depend on the degree of correlation between the $t$ and $\bar t$
spins, and hence on other variables such as new production mechanisms,
center of mass energy, parton
distributions, or the presence of
anomalous $gtt$ couplings. This dependence on the production process makes the
analysis of the $Wtb$ vertex with spin correlations less clean
than with the FB asymmetry.

\section{Dependence of the FB asymmetry on couplings and masses}

\begin{figure}[htb]
\begin{center}
\mbox{\epsfig{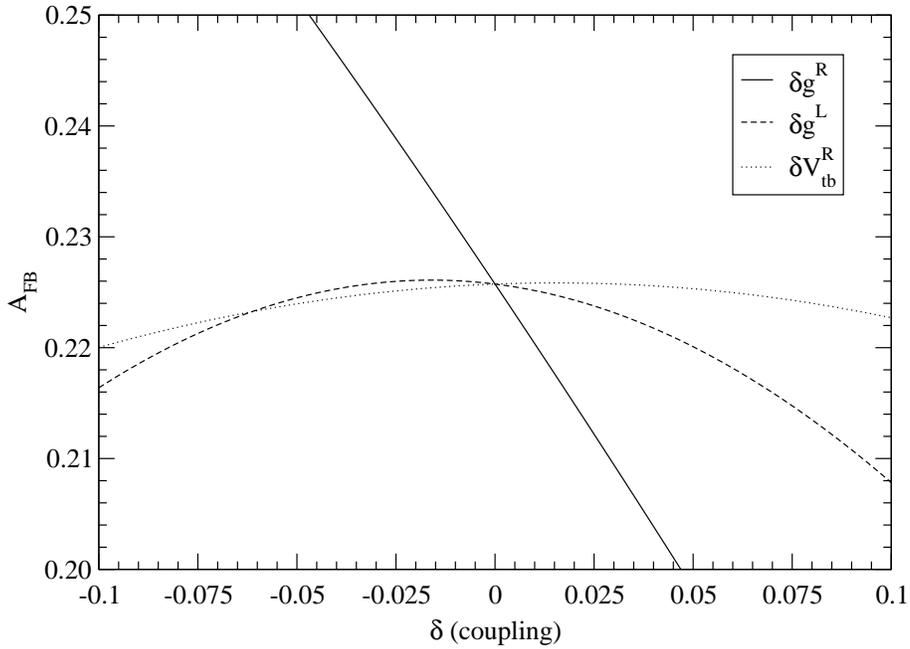}}
\end{center}
\caption{Dependence of $A_\mathrm{FB}$ on $\delta g^R$ (solid line),
$\delta g^L$ (dashed line) and $\delta V_{tb}^R$ (dotted line) using the
analytical expression.
\label{fig:afb} }
\end{figure}

Before the numerical discussion it is enlightening to have a look at the
dependence of $A_\mathrm{FB}$ (which will be the best observable among the
asymmetries discussed) on the nonstandard
couplings. For simplicity we ignore corrections from the $t$ and $W$ widths,
which will turn out to be irrelevant, but keep $m_b$ nonvanishing. Unlike
$A_{1-3}$, $A_\mathrm{FB}$ only depends on the $t$, $b$ and $W$ boson masses,
and on the couplings in Eq.~(\ref{ec:1}). For small $g^R$ the leading dependence
of $A_\mathrm{FB}$ on this coupling is given by the interference term
$V_{tb}^L g^R$.
For $g^R = 0$ we obtain the SM tree-level (LO) value $A_\mathrm{FB} = 0.2223$. 
The bulk effect of one-loop QCD corrections in the asymmetry can be taken into
account including a $\sigma^{\mu \nu}$ term $g^R = -0.00642$ \cite{papiro8}.
The corresponding NLO value is $A_\mathrm{FB} = 0.2257$. Borrowing from our
numerical analysis below the combined statistical error of
the asymmetry, $\delta A_\mathrm{FB} \simeq 5 \times 10^{-4}$,
we obtain the simple estimate that QCD corrections amount to a $10\sigma$
effect. This high sensitivity is due to the presence of the large linear
term. The dependence of $A_\mathrm{FB}$ on $g^L$ and $V_{tb}^R$ is mainly
quadratic because the linear terms are suppressed. Then, the sensitivity to
these couplings is less impressive.
In Fig.~\ref{fig:afb} we plot $A_\mathrm{FB}$
for different values of $\delta g^R \equiv g^R +
0.00642$, $\delta g^L \equiv g^L$ and $\delta V_{tb}^R \equiv V_{tb}^R$. 
(We use $m_t = 175$, $M_W =
80.33$, $m_b = 4.8$ GeV.)

\begin{figure}[htb]
\begin{center}
\mbox{\epsfig{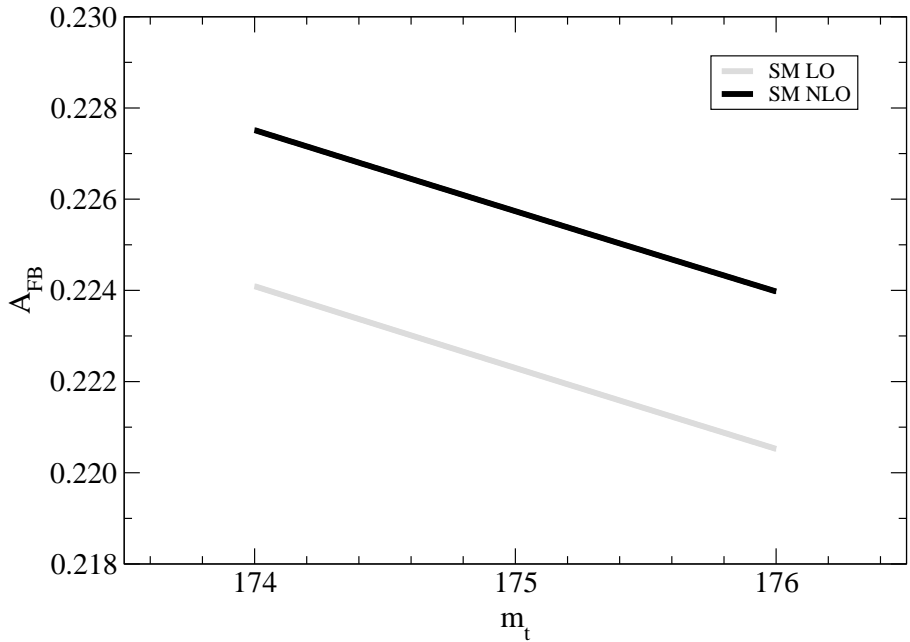}}
\end{center}
\caption{Dependence of $A_\mathrm{FB}$ on the top mass for $m_t = 175 \pm 1$
GeV. The thickness of the lines shows the variation for
$M_W = 80.33 \pm 0.015$ GeV.
\label{fig:afb2} }
\end{figure}

We expect that the main source of systematic error on $A_\mathrm{FB}$ will be
the uncertainty in the top mass. For $\delta m_t \simeq 1$ GeV, as estimated
for LHC, the variations in the LO and NLO predictions of $A_\mathrm{FB}$ with
respect to the central values with $m_t = 175$ GeV are shown in
Fig.~\ref{fig:afb2}. The thickness of the lines corresponds to the
uncertainty in the $W$ mass, which is expected to be measured at LHC with a
precision of $0.015$ GeV, and has a much smaller effect on the FB asymmetry.
The full potential of the measurement of $A_\mathrm{FB}$ will be available
when $\delta m_t$ is reduced to $0.15$ GeV and $\delta M_W$ to $0.006$ GeV at
TESLA (see Ref.~\cite{papiro22} and references there in). With these precisions,
the systematic errors due to $m_t$ and $M_W$ are almost negligible.

\section{Numerical results}

The calculation of the matrix elements for the $2 \to 6$ processes
$gg,\;q \bar q \to t \bar t \to W^+ b W^- \bar b \to l \nu jjjj$,
including all spin correlations,
is performed using HELAS extended to compute the $\sigma^{\mu \nu}$
vertices present in the top and antitop decays. As background we
consider $W^\pm$ plus four jets, calculated with VECBOS \cite{papiro10}. We use
the MRST structure functions set A \cite{papiro11} with $Q^2 = \hat s$.
We generate enough events to ensure that
the Monte Carlo uncertainties are below the experimental statistical
errors, and use the same event sets (one for the $\mu$ channel and another
for the $e$ channel) for all the evaluations of the asymmetries.
This is crucial to make certain that the small deviations in the
asymmetries are not an artifact of the numerical integration.

Assuming for the moment perfect particle identification and neglecting any
detector effects, we obtain for the FB asymmetry the
LO value $A_\mathrm{FB} = 0.2222$ and the NLO value $A_\mathrm{FB}=0.2257$. These
are remarkably close to the ones obtained above neglecting corrections from the
top and $W$ widths. These corrections
cancel after integration in phase space due to their linear dependence on
the small ratios $\Gamma_t/m_t$ and $\Gamma_W/M_W$.
Note also that
NLO corrections to the production of $t \bar t$ pairs do not modify the FB
asymmetry, and can be taken into account with a global factor $K = 1.7$
\cite{papiro1}. For the spin correlation asymmetry $A_1$
we obtain the tree-level value $-0.0835$, in agreement with
Ref.~\cite{papiro14} within Monte Carlo uncertainties.
The sensitivity of the asymmetries $A_{1-3}$ to anomalous couplings is not
modified by NLO corrections to $t \bar t$ production, which
have little influence on top-antitop spin correlations at LHC
\cite{papiro18}. These corrections must be included when
comparing with real data. In this case it is also compulsory to take
hadronisation into account and to use a proper detector simulation.
In order to estimate the
sensitivity to anomalous $Wtb$ couplings, it is sufficient
to simulate the calorimeter resolution by performing a Gaussian
smearing of the energies. We use
\begin{eqnarray}
\frac{\Delta E^j}{E^j} & = & \frac{50\%}{\sqrt{E^j}} \oplus 3\% \,, \nonumber \\
\frac{\Delta E^e}{E^e} & = & \frac{10\%}{\sqrt{E^e}} \oplus 0.3\% \,,
\nonumber \\
\frac{\Delta E^\mu}{E^\mu} & = & 2.5\%
\label{ec:6b}
\end{eqnarray}
for jets, electrons and muons, respectively.
The energies in Eqs.~(\ref{ec:6b}) are GeV and the terms are added in
quadrature. We apply ``detector'' cuts on transverse momenta of the jets
$p_T^j \geq 20$ GeV, electrons $p_T^e \geq 15$ GeV, and muons $p_T^\mu \geq 6$
GeV. We also require pseudorapidities $|\eta| \leq 2.5$ and reject the events
where the charged lepton and/or the jets are not isolated, requiring a distance
in ($\eta$, $\phi$) space $\Delta R \geq 0.4$.
We then require the signal and background events to fulfill at least one of the
ATLAS trigger conditions. In practice, at the high luminosity phase
they imply the cuts $p_T^e \geq 30$ GeV in the
electron channel and $p_T^\mu \geq 20$ GeV in the muon channel. These
conditions reduce the phase space in the forward region
and then $A_\mathrm{FB}$ to $\sim 0.15$ for muons and $\sim 0.05$ for
electrons. For $A_{1,2}$, the effect of the detector simulation
is to reduce them by factors of $0.55$, $0.7$ in the muon channel and
$0.15$, $0.4$ in the electron channel, respectively.
$A_3$ is washed out in both channels.

The events are reconstructed identifying first the three jets from the antitop
hadronic decay.Among the four final jets, two of them $j_1$, $j_2$ must
reproduce the $W$ mass $(M_W^\mathrm{rec})^2 = (p_{j_1}+p_{j_2})^2$, and with a
third one $j_3$ the $\bar t$ mass 
$(m_{\bar t}^\mathrm{rec})^2 = (p_{j_1}+p_{j_2}+p_{j_3})^2$. Of the twelve
possible
combinations, we choose the one minimising the sum of square mass differences
$(m_{\bar t}^\mathrm{rec}-m_t)^2+(M_W^\mathrm{rec}-M_W)^2$.
The remaining jet is then assigned to the $b$
quark. With this kinematic identification $b$ tagging is neither necessary
nor convenient in order to keep the signal as large as possible. The neutrino
momentum $p_\nu$ is reconstructed from the missing transverse momentum
$p_T\!\!\!\!\!\!\!\!\! \not \,\,\,\,\,\,\,\,$ and the charged lepton momentum
$p_l$, identifying
$(p_\nu)_T = p_T\!\!\!\!\!\!\!\!\! \not \,\,\,\,\,\,\,\,$ and solving
$(p_\nu + p_l)^2 = M_W^2$ for $(p_\nu)_L$. Of the two possible values for the
longitudinal
momentum we choose the solution with $(p_\nu + p_l + p_b)^2$ closer to
$m_t^2$, what ensures the correct event reconstruction.

The presence of a calculable $W\!jjjj$ background does not change
the value of the asymmetries but increases their statistical error.
Thus, it is convenient to reduce the background as much as possible
without spoiling
the signal. For this purpose we apply loose cuts on the reconstructed masses
of $t$, $\bar t$ and $W^-$,
\begin{eqnarray}
130 \leq & \hspace*{-0.1cm} m_t^\mathrm{rec} \hspace*{-0.1cm}
  & \leq 220 \,, \nonumber \\
150 \leq & \hspace*{-0.1cm} m_{\bar t}^\mathrm{rec} \hspace*{-0.1cm}
  & \leq 200 \,, \nonumber \\
65  \leq & \hspace*{-0.1cm} M_{W^-}^\mathrm{rec} \hspace*{-0.1cm}
  & \leq 95 \,,
\end{eqnarray}
which have very little effect on the signal. 
The mass window is wider for $m_t^\mathrm{rec}$ than for 
$m_{\bar t}^\mathrm{rec}$ because the reconstruction of
the top from missing transverse momentum is worse than the antitop
reconstruction from three jets.
In Table~\ref{tab:1} we collect
the signal and background cross-sections in the SM.
The statistical errors for the asymmetries computed from this Table
are $\delta A \simeq 6.5 \times 10^{-4}$ and
$\delta A \simeq 7.3 \times 10^{-4}$
in the muon and electron channels, respectively, for an integrated luminosity
of 100 fb$^{-1}$.
In Table~\ref{tab:2} we collect the asymmetries
within the SM and for several values of the anomalous parameters.
Using these figures the statistical significance of the deviations 
from the SM NLO prediction can be
computed for each channel and combined to yield the total statistical
significance in Table~\ref{tab:3}.
Notice that although the asymmetries for the $e$ channel are typically smaller
than for the $\mu$ channel, the sensitivities are similar.
We also observe that the significances obtained are 60--70\% of those
obtained naively from Figure~\ref{fig:afb} and the combined statistical error.

\begin{table}[p]
\begin{center}
\begin{tabular}{llcccc}
\hline
\hline
 & & \multicolumn{2}{c}{$l=\mu$} & \multicolumn{2}{c}{$l=e$} \\
 & & $t \bar t$ & $W\!jjjj$ & $t \bar t$ & $W\!jjjj$ \\
\hline
$A_\mathrm{FB}$ & $N_F$ & 16.065 & 2.62 & 12.227 & 2.99 \\
       & $N_B$ & 11.874 & 2.59 & 11.009 & 2.42 \\
$A_1$  & $N_F$ & 13.641 & 2.51 & 11.544 & 2.66 \\
       & $N_B$ & 14.297 & 2.71 & 11.693 & 2.74 \\
$A_2$  & $N_F$ & 14.523 & 2.82 & 11.907 & 2.94 \\
       & $N_B$ & 13.415 & 2.40 & 11.330 & 2.46 \\
$A_3$  & $N_F$ & 13.956 & 2.62 & 11.550 & 2.67 \\
       & $N_B$ & 13.983 & 2.59 & 11.687 & 2.73 \\
\hline
\hline
\end{tabular}
\caption{Signal and background cross-sections (in pb) in the F and B hemispheres
for each asymmetry and decay channel.
\label{tab:1}}
\end{center}
\end{table}

\begin{table}[p]
\begin{center}
\begin{tabular}{llcccc}
\hline
\hline
Coupling              & $l$    & $A_\mathrm{FB}$ & $A_1$ & $A_2$ & $A_3$ \\
\hline
SM (NLO)                & $\mu$ & 0.1500 & -0.0235 & 0.0397 & -0.0010 \\
SM (NLO)                & $e$   & 0.0524 & -0.0064 & 0.0248 & -0.0059 \\
SM (LO)                 & $\mu$ & 0.1470 & -0.0236 & 0.0398 & -0.0012 \\
SM (LO)                 & $e$   & 0.0496 & -0.0066 & 0.0250 & -0.0061 \\
$\delta g^R = +0.003$    & $\mu$ & 0.1486 & -0.0235 & 0.0398 & -0.0011 \\
$\delta g^R = +0.003$    & $e$   & 0.0511 & -0.0065 & 0.0249 & -0.0060 \\
$\delta g^R = -0.003$   & $\mu$ & 0.1514 & -0.0234 & 0.0396 & -0.0009 \\
$\delta g^R = -0.003$   & $e$   & 0.0537 & -0.0063 & 0.0247 & -0.0058 \\
$\delta g^L = +0.02$     & $\mu$ & 0.1486 & -0.0233 & 0.0396 & -0.0010 \\
$\delta g^L = +0.02$     & $e$   & 0.0510 & -0.0064 & 0.0247 & -0.0059 \\
$\delta g^L = -0.05$    & $\mu$ & 0.1488 & -0.0234 & 0.0396 & -0.0010 \\
$\delta g^L = -0.05$    & $e$   & 0.0512 & -0.0064 & 0.0247 & -0.0059 \\
$\delta V_{tb}^R = +0.08$  & $\mu$ & 0.1485 & -0.0229 & 0.0390 & -0.0010 \\
$\delta V_{tb}^R = +0.08$  & $e$   & 0.0510 & -0.0061 & 0.0242 & -0.0059 \\
$\delta V_{tb}^R = -0.04$ & $\mu$ & 0.1488 & -0.0231 & 0.0393 & -0.0009 \\
$\delta V_{tb}^R = -0.04$ & $e$   & 0.0512 & -0.0062 & 0.0245 & -0.0058 \\
\hline
\hline
\end{tabular}
\caption{Asymmetries for some representative values of the anomalous couplings.
\label{tab:2}}
\end{center}
\end{table}

\begin{table}[ht]
\begin{center}
\begin{tabular}{lcccc}
\hline
\hline
Coupling                & $A_\mathrm{FB}$ & $A_1$ & $A_2$ &  $A_3$ \\
\hline
SM (NLO)             & $6.1\sigma$ & $0.3\sigma$ & $0.4\sigma$ & $0.4\sigma$ \\
$\delta g^R=+0.003$  & $2.8\sigma$ & $0.1\sigma$ & $0.2\sigma$ & $0.2\sigma$ \\
$\delta g^R=-0.003$  & $2.8\sigma$ & $0.1\sigma$ & $0.2\sigma$ & $0.2\sigma$ \\
$\delta g^L=+0.02$   & $3.0\sigma$ & $0.3\sigma$ & $0.1\sigma$ & $0.0\sigma$ \\
$\delta g^L=-0.05$  & $2.5\sigma$ & $0.1\sigma$ & $0.1\sigma$ & $0.0\sigma$ \\
$\delta V_{tb}^R=+0.08$ & $3.1\sigma$ & $0.9\sigma$ & $1.2\sigma$ & $0.0\sigma$
\\
$\delta V_{tb}^R=-0.04$ & $2.6\sigma$ & $0.6\sigma$ & $0.7\sigma$ & $0.1\sigma$
\\
\hline
\hline
\end{tabular}
\caption{Combined statistical significance of the deviations in the asymmetries
in Table \ref{tab:2}.
\label{tab:3}}
\end{center}
\end{table}

We have ignored systematic errors in our study. As we have mentioned in the
previous Section, we expect that the main source of systematic error will be the
uncertainty in the top mass (see Fig.~\ref{fig:afb2}). We have given in
Table~\ref{tab:3} the precision in the determination of anomalous couplings
that will be possible when TESLA reduces the uncertainty in the top mass to
$\delta m_t = 0.15$ GeV. Before TESLA operation, the theoretical uncertainty due
to $\delta m_t$ must be taken into account in each decay channel.
The bulk effect of this systematic uncertainty for the expected LHC precision
$\delta m_t = 1$ GeV is that the figures in Table~\ref{tab:3} have to be
reduced by factors of $0.6-0.7$, being the difference between the SM LO and NLO
predictions approximately $3.6 \sigma$ in this case, and the remaining
statistical significances between $1.8\sigma$ and $2 \sigma$.
Additionally, it should be pointed out
that if we assumed that no new physics contributes to the
$Wtb$ vertex, the measurement of the FB asymmetry could be turned into an
indirect, model-dependent determination of the top mass
\cite{papiro19} with an accuracy of
$\delta m_t \simeq 0.5$ GeV.

The FB asymmetry will be first observed at Tevatron, but the small statistics
available will not allow to perform precision tests. With a similar analysis
and the cuts $p_T^{e,\mu,j} \geq 10$ GeV,
$|\eta^e| \leq 2$, $|\eta^\mu| \leq 1.5$, $|\eta^j| \leq 2.5$, we obtain for
an integrated luminosity of 2 fb$^{-1}$
$A_\mathrm{FB} = 0.21 \pm 0.04$ and $A_\mathrm{FB} = 0.23 \pm 0.08$ in the $e$
and $\mu$ channels,
respectively. The asymmetry can be measured with $5.4\sigma$, but
the anomalous couplings needed to have a $3\sigma$ deviation
are large, $|g_{R,L}| \sim 0.3$, $|V_{tb}^R| \sim 0.7$.

In summary, we conclude that $A_\mathrm{FB}$ is an excellent tool for the study
of the $Wtb$ vertex that can spot one-loop QCD corrections with $6\sigma$
significance. The sensitivity to $g^R$ is one order of magnitude better than in
single top production at LHC \cite{papiro7}, and even at a 1000 GeV
$\gamma e^-$ collider with a luminosity of 500 fb$^{-1}$ \cite{papiro20}. The
sensitivity to $g^L$ is similar but better than the one expected at a linear
$e^+ e^-$ or $\gamma e^-$ collider, or in single top production at LHC. Spin
correlation asymmetries are not quite as sensitive and they depend on the
production process as well. This fact has, however, a bonus: if $A_\mathrm{FB}$
has its predicted value and $A_1$ or $A_2$ do not, then the source of the
discrepancy is bound to be an anomalous coupling or mechanism in $t \bar t$
production. Hence, the study of the FB asymmetry in $t\bar t$ production also
complements spin correlation asymmetries helping to disentangle the origin of
new physics, if observed.

\vspace{1cm}
\noindent
{\Large \bf Acknowledgements}

\vspace{0.4cm} \noindent
This work has been supported by the European Community's Human Potential
Programme under contract HTRN--CT--2000--00149 Physics at Colliders and by MCYT
and Junta de Andaluc\'{\i}a.

\end{document}